# Optical power detector with broad spectral coverage, high detectivity and large dynamic range

Jussi Rossi[1], Juho Uotila[2], Sucheta Sharma[3], Tuomas Hieta[4], Toni Laurila[3], Roland Teissier[5], Alexei Baranov[5], Erkki Ikonen[3,6] and Markku Vainio[1,7]

1. Photonics Laboratory, Physics Unit, Tampere University, Tampere, Finland
2. Patria Aviation Oy, Tampere, Finland
3. Metrology Research Institute, Aalto University, Espoo, Finland
4. Gasera Ltd, Turku, Finland
5. IES, University of Montpellier, CNRS, Montpellier, France
6. VTT MIKES, Espoo, Finland
7. Department of Chemistry, University of Helsinki, Helsinki, Finland

Correspondence: Markku Vainio (markku.vainio@helsinki.fi)

**Abstract:** Optical power measurements are needed in practically all technologies based on light. Here we report a general-purpose optical power detector based on the photoacoustic effect. Optical power incident on the detector's black absorber produces an acoustic signal, which is further converted into an electrical signal using a silicon-cantilever pressure transducer. We demonstrate an exceptionally large spectral coverage from ultraviolet to far infrared, with the possibility for further extension to the terahertz region. The linear dynamic range of the detector reaches 80 dB, ranging from a noise-equivalent power of 6 $nW/\sqrt{Hz}$ to 600 mW.

## Introduction

Power detector is one of the most important components in technologies based on light or other electromagnetic radiation. Depending on the application, typical detector parameters of interest include noise-equivalent power (NEP), linear dynamic range, speed, spectral coverage, and ease of use. Semiconductor photodetectors, such as Si photodiodes, provide excellent performance with respect to many of these parameters, but the spectral coverage is limited on the long-wavelength side by the bandgap energy of the detector material. Although extension to long wavelengths is in some cases possible by material engineering, the reduced bandgap inevitably leads to increased thermal noise and thus to the need for cooling of the detector, even down to cryogenic temperatures [1]. In practice, the responsivity of a typical state-of-the-art HgCdTe (MCT) infrared detector sharply rolls off at about 12 μm [2]. Beyond that, thermal detectors are often used. In a thermal detector, a black material absorbs the incoming optical radiation and the resulting temperature increase of the material is measured [2]. An advantage common to all thermal detectors is that they can be made practically wavelength independent [3]. This allows access to the long infrared part of the electromagnetic spectrum, which is becoming increasingly important due to numerous applications based on molecular spectroscopy and thermal imaging – a trend that is catalyzed by the fast development of coherent infrared light sources [4-8].

The most common thermal detectors include bolometers, pyroelectric detectors and thermopiles, all of which utilize temperature dependence of an electric property of the detector material [2,9]. The photoacoustic (PA) effect offers an alternative approach [10]. In the PA effect, the temperature increase due to absorption of the incident electromagnetic radiation induces a pressure change in the surrounding gas. If the incident radiation is modulated, similar to what is done with *e.g.* pyroelectric detectors, the temperature change and the corresponding pressure change become periodic (Fig. 1a). That is, an acoustic signal directly proportional to the absorbed power is produced. The acoustic signal can be converted into an electrical signal by a sensitive microphone.

The photoacoustic detection principle is versatile and tailorable for various applications. As an example, ultrasound transducers that operate at MHz frequencies provide high spatial resolution in fast photoacoustic imaging in medical applications [11-13]. On the other hand, some of the best detection limits in gas-phase molecular spectroscopy have been obtained by optimizing the detection at lower (10 Hz to 100 kHz) acoustic frequencies [14,15], an approach that has recently opened new avenues also in broadband laser spectroscopy [16,17]. The use of the PA effect in optical power detection is known from the 1940's, when Golay developed a photoacoustic cell particularly designed for this purpose [18]. The Golay cell is one of the best thermal detectors in terms of detection sensitivity, but its diaphragm microphone is fragile and susceptible to technical noise caused by mechanical vibrations [9]. The dynamic range of the diaphragm microphone is also modest, limiting the measurable optical power range to approximately 100 nW to 1 mW for an optimized commercially available instrument [19]. The measurement range can be shifted towards lower powers, and NEP values of close to or below $1\ \mathrm{nW}/\sqrt{\mathrm{Hz}}$ have been reported for both Golay cells and miniaturized versions based on tunneling displacement transducers [9,20,21]. The poor linearity of the diaphragm microphone poses an additional challenge [22,23]. Due to these limitations, the Golay cell mainly finds use in niche laboratory applications, although the fundamental underlying principle, photoacoustic detection, offers the possibility for one of the most versatile high-performance thermal detectors.

In this article, we introduce a photoacoustic power detector that has an excellent dynamic range combined with a rugged construction. Our solution is based on cantilever-enhanced PA detection, which is one of the most sensitive methods developed for optical gas analysis [14,24,25]. Here, we show that the same principle can lead to a simple and universal electromagnetic power detector that works virtually at any wavelength without cooling, thus benefiting a wide range of applications. While the main motivation of our work is to develop a general-purpose optical power detection method, the work also contributes to the development of traceable power measurements in the long-wavelength infrared and terahertz (THz) regions, where power calibrations are particularly difficult to carry out. An elegant solution proposed for reduction of the uncertainty of such calibrations is to link the long-wavelength power responsivity scale to a more accurate responsivity scale in the visible spectral range [26]. This requires a reference detector with broad spectral coverage and predictable spectral responsivity.

## Results

The operating principle of the cantilever-enhanced power detector is shown schematically in Fig. 1a. The radiation incident on the detector is first modulated with a mechanical chopper and then focused on the absorber inside of a sealed photoacoustic cell (see "Methods" for details of the components). Due to the large diameter ($\geq$ 10 mm) of both the cell window and the absorber, the detector is also suitable for measurements of incoherent radiation and long wavelengths. The PA cell is filled with an acoustic carrier gas, typically air, $N_2$ or He [27,28]. The acoustic wave generated by modulated absorption is detected by a silicon cantilever [24]. It is mechanically robust and suitable for field applications despite being only a few micrometers thick [29]. The cantilever deflection is measured at picometer resolution using a compact laser-interferometric readout module that produces an electrical signal directly proportional to the absorbed optical power [30]. The signal is recorded by monitoring the signal peak in the photoacoustic frequency spectrum (Fig. 1b) that is obtained from the digitized interferometer signal in real time by fast Fourier transform (FFT).

Unlike with many other PA sensors [11,15], the acoustic frequency of cantilever-enhanced PA detection can be chosen freely; there is no need to match the modulation frequency with a narrow acoustic resonance peak of the cell. This property simplifies the detector design and, as is shown below, gives an additional method to increase the dynamic range of PA detection. With the cantilever used in this work, the best signal-to-noise ratio (SNR) is obtained at modulation frequencies of the order of 40 Hz [27].

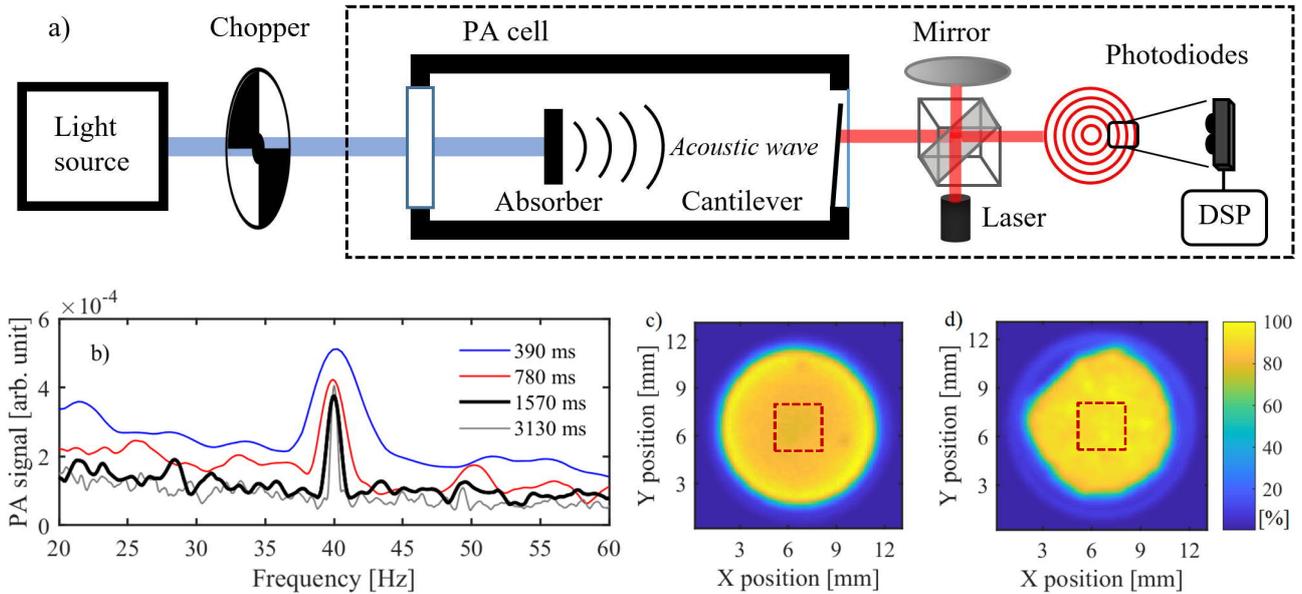

**Fig. 1. Working principle of the cantilever-enhanced PA power detector. a** Simplified experimental setup. The dashed box shows the power detector components and DSP = digital signal processing unit. **b** Examples of PA spectra obtained by FFT of the cantilever signal. The spectrum shows the PA noise floor and, on top of that, the actual signal at the modulation frequency, which in this case is 40 Hz. Signal peaks were recorded with different measurement times (FFT time constants) and with 50 nW of incident optical power at λ = 633 nm. **c** Spatial PA response of the soot absorber and **d** of the CNT absorber. The dashed-line red squares show the 3 mm x 3 mm areas used in spatial uniformity calculations, see "Methods" for details.

The proof-of-concept study reported here covers a wide range of wavelengths, from 325 nm to 25 µm. Based on rigorous characterization of the photoacoustic properties of several black absorber materials [27], we have chosen a simple absorber fabricated in-house by depositing candle soot on an aluminum substrate, as well as a commercially available carbon nanotube (CNT) sample. The soot-based absorber gives a larger PA signal while the CNT absorber has a flatter spectral response; see Fig. 2a. The dashed-line plot in Fig. 2b was measured with a broadband thermal light source, while the individual dots summarize the results of laser-based calibrations at several wavelengths, from 325 nm to 14.85 µm (see "Methods").

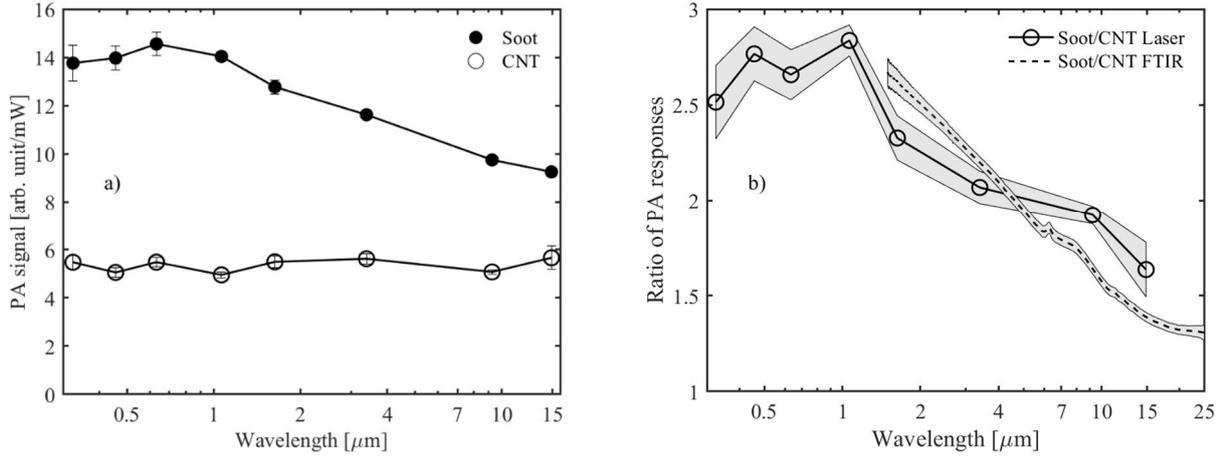

**Fig. 2. Spectral responsivity of the PA detector. a** The PA response of the detector with two different absorbers measured at discrete laser wavelengths. **b** Comparison of the two absorbers, soot on aluminum and a CNT sample. The dashed line between wavelengths 1.5 and 25 µm was measured with an incandescent light source and a Fourier-transform infrared (FTIR) spectrometer. The individual data points from 325 nm to 14.85 µm were recorded with various laser sources. The slight discrepancy between these two datasets is due to a wavelength-dependent artifact in the FTIR measurements (see "Methods").

One of the advantages of the cantilever microphone is its excellent linear dynamic range. The linearity of a conventional microphone is limited by the nonlinear stretching of the diaphragm. Additional nonlinear contribution arises from the electrostatic forces if capacitive readout of the microphone signal is used. The silicon cantilever is attached to its frame from only one side, which results into a highly linear motion that is precisely monitored by the interferometric readout unit. The large linear dynamic range of 60 dB is exemplified in Fig. 3a. With the soot-based absorber and He as the acoustic carrier gas, the normalized NEP measured at 1064 nm is 5.6 $nW/\sqrt{Hz}$, corresponding to a specific detectivity of $D^* = 1.6 \times 10^8$ $cm\sqrt{Hz}/W$. These values compare favorably to state-of-the-art uncooled mid-infrared detectors, such as pyroelectric detectors coated with carbon nanotubes [31,32]. The upper limit (8.2 mW at 40 Hz modulation frequency) of the dynamic range is in this case set by the readout interferometer. This linear dynamic range is at least 20 dB better than that of a typical MCT detector [33] or Golay cell [22] and can be further enhanced by varying the modulation frequency, because the photoacoustic response drops as the modulation frequency is increased [27]. Adjustment of the chopping frequency and/or chopping duty cycle allows us to extend the maximum measurable power on the fly to over 600 mW without saturating the detector (Fig. 3b). As a result, we achieve a total linear dynamic range of 80 dB with a detection bandwidth of 1 Hz (see "Methods" for the definition and calculation of the dynamic range).

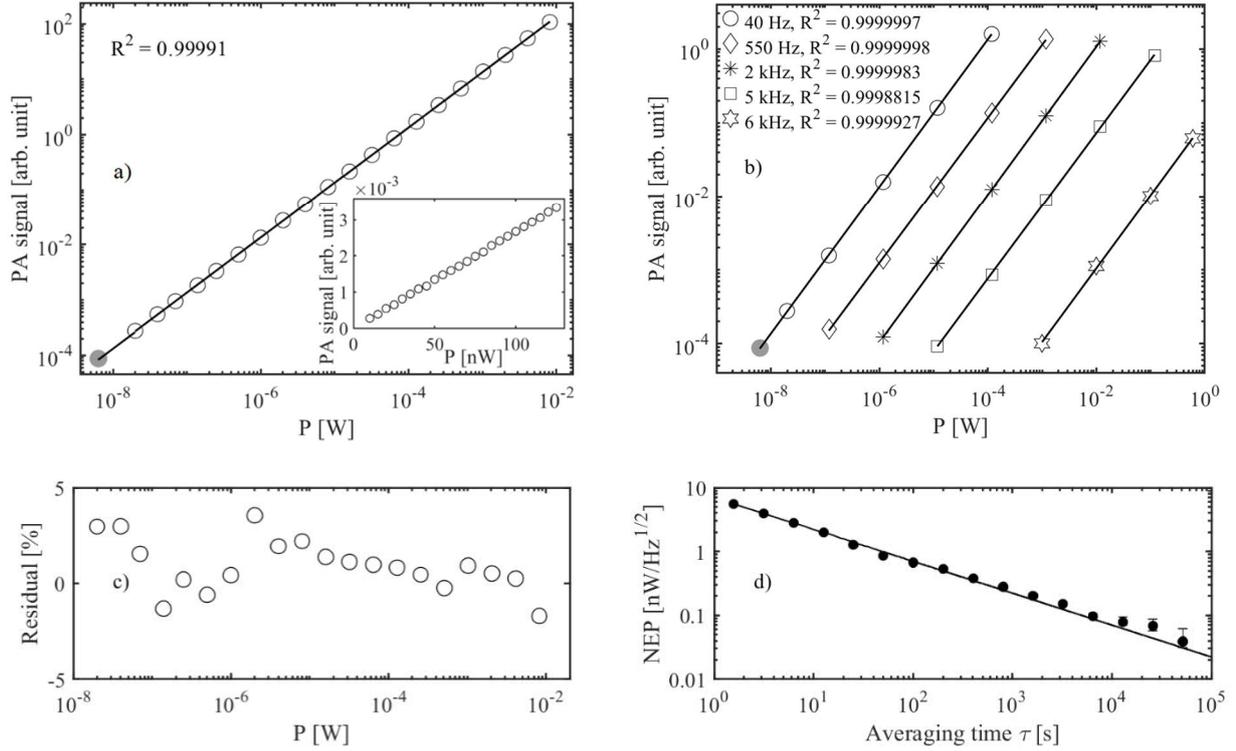

**Fig. 3. Linearity and NEP of the photoacoustic power detector using soot-based absorber. a** The circles indicate the measured PA signal as a function of incident optical power at 1064 nm wavelength, measured with 1.57 s averaging time and 40 Hz modulation frequency. Solid dot indicates the noise-equivalent power at 40 Hz modulation frequency and the solid line is a linear fit to the data. The inset shows low-power points on a linear scale. **b** Extension of the linear dynamic range by adjustment of the chopping frequency up to 6 kHz. **c** The relative residual of plot 3a, *i.e.* the relative deviation of the measured points from the linear fit. **d** Dependency of the detector NEP on the averaging time $\tau$, derived from Allan-deviation analysis (see "Methods"). The straight line indicates the white-noise slope $\tau^{-1/2}$ of Allan deviation.

As mentioned above, PA detection can be designed for different data acquisition speeds. General power measurements tend to require a low detection limit but not a fast response time, for which reason we have designed the device for low modulation frequencies. We typically use a measurement time (FFT time constant) of 1.57 s, although shorter times are possible, as illustrated in Fig. 1b. On the other hand, some applications can use even longer averaging times, benefitting from the excellent stability and white-noise limited performance of the cantilever microphone (Fig. 3d).

## Conclusions

In summary, we have demonstrated a next-generation photoacoustic power detector that is based on a robust silicon-cantilever microphone, is easy to use and works without cooling. Our proof-of-concept device works from ultraviolet to long infrared wavelengths, significantly exceeding the spectral coverage of semiconductor photodetectors. Further extension to the THz region is possible. Together with the excellent linear dynamic range, the wavelength-independent concept is appealing not only for general purposes but also for the development of power meters and transfer standards that can reduce the uncertainty of long-wavelength power measurements (given that rigorous metrological characterization of the detector is first done over the entire spectral range). The strength of the PA signal scales favorably with smaller acoustic cell size [30], which together with the compact cantilever microphone offers room for substantial miniaturization of the instrument. On the other hand, the concept is ideal for the detection of thermal light and long-wavelength radiation because the NEP is independent of the absorber size.

## Materials and methods

Photoacoustic sensor

The photoacoustic sensor is based on a commercially available instrument, PA301 (Gasera Ltd). The photoacoustic cell (absorber cell volume 0.4 cm$^2$ and total gas volume 7.6 cm$^2$) of the instrument is equipped with gas valves, which allowed us to test various acoustic carrier gases [27,28]. Here, helium was used in all experiments. The absorber and window aperture are 10 mm and 15 mm in diameter, respectively. We used a 2 mm thick KBr window, which has a high transmittance within the entire wavelength range covered (0.3 to 25 µm). The soot-based absorber was fabricated by using a candle flame to deposit a 0.5-mm thick layer of soot on the 0.5-mm thick bottom of an Al cup [27]. The CNT-based absorber has a spray-applied coating of carbon nanotubes (Vanta S-VIS, Surrey NanoSystems Ltd) on aluminum foil, which was cut to a nearly circular form (Fig. 1d). The measured spatial uniformities of the absorbers' PA responses are shown in Figs. 1c-d. The measurement technique for spatial uniformity is described in reference [27]. The standard deviation describing spatial variations of the PA response, as calculated for the 3 mm x 3 mm areas highlighted in Figs. 1c-d, is less than 3 % for both the soot and CNT absorber. This value is similar to those reported for CNT-coated pyroelectric detectors [31,32]. Note that the spatial uniformity at the center of absorber is equally good for soot and CNT; the uniformity of the soot sample (Fig. 1c) just seems to be better than that of the CNT disk (Fig. 1d) due to the strongly absorbing ring of the soot sample, which downshifts the absolute readings of the center area on the scale bar. This ring is due to the sooting process, which accumulates soot on the edges of the metal cup substrate [27,28].

The silicon cantilever was microfabricated by a silicon-on-insulator wafer etching process [29] and installed in the photoacoustic cell at 13 mm distance from the absorber center. The dimensions of the cantilever used here are 4 mm x 1 mm x 10 µm (L x W x T). The cantilever displacement due to an acoustic wave is recorded with a compact laser interferometer that is integrated in the PA301 instrument. The digitized output signal of the interferometer module is processed in a LabVIEW program to perform fast Fourier transform (FFT) of the PA signal. The signal is measured by monitoring the value of the FFT peak at the modulation frequency (see Fig. 1b). Alternatively, phase-sensitive measurement of the interferometer signal is possible using a lock-in amplifier referenced to the chopper signal. Phase-sensitive detection with a lock-in amplifier can improve the PA-detection SNR by a factor of 2 compared to the FFT method [34]. However, in the proof-of-concept study reported here, we used the FFT method to retrieve full information of the photoacoustic noise spectrum.

Measurements

The photoacoustic power detector was characterized using several monochromatic continuous-wave lasers: 325 nm HeCd laser, 457 nm Ar-ion laser, 633 nm HeNe laser, 1064 nm amplified Yb-fiber laser, 1.63 µm external-cavity diode laser, 3.39 µm HeNe laser, 9.24 µm quantum cascade laser and a 14.85 µm quantum cascade laser [6]. In each case, the power of the collimated laser beam was modulated in transmission mode using a mechanical chopper with a variable chopping frequency. We typically used a 50 % duty cycle and 40 Hz chopping frequency, but Fig. 3b also shows how the dynamic range can be extended by increasing the chopping frequency. The last data set of Fig. 3b was measured at 6 kHz chopping frequency and 20 % duty cycle. The smaller duty cycle was used to avoid any nonlinearity at high chopping frequencies arising from the detector's thermal recovery time constant. Moreover, the small duty cycle reduces the average optical power incident on the detector and thus minimizes the risk of power-induced damage of the absorber.

The laser beam was aligned to the center of the absorber through several irises. The laser spot diameter on the absorber was smaller than 3 mm in all cases, so that the beams always hit the spatially uniform regions highlighted by the red squares in Figs. 1c-d. Furthermore, we have experimentally confirmed that the PA response does not depend on the laser spot diameter when the diameter is in the range of 1 to 3 mm. Optical power adjustment was done either directly by changing the laser drive current (1.63 µm, 9.24 µm and 14.85 µm) or using a variable neutral-density filter (325 nm, 457 nm, 633 nm and 1064 nm). Each laser measurement was done by comparing the PA signal to the signal of a reference detector that was calibrated for SI-traceability; a NIST-calibrated silicon detector or a thermal detector that was calibrated using a pyroelectric radiometer of the Metrology Research Institute (Aalto University, Finland) [35].

The continuous broadband infrared measurements that cover wavelengths from 1.5 µm to 25 µm (Fig. 2b) were carried out using a SiC incandescent light source. The light was first passed through a Fourier-transform infrared (FTIR) spectrometer (Bruker IRCube Matrix M series), which interferometrically modulates the light power such that each optical frequency $\nu_{IR}$ is mapped to an acoustic frequency $\nu_{PA} = \nu_{IR} c_{FTIR}$, where $c_{FTIR} \sim 3.38 \times 10^{-12}$ is the down-conversion factor that depends on the FTIR scan speed (0.1013 cm/s optical path length change). The PA detector measures the respective FTIR interferogram, from which the photoacoustic spectrum is obtained by FFT as described above. The optical spectrum is reconstructed from the PA spectrum by multiplying its frequency scale with the factor $c_{FTIR}$. It is important to note that the effective modulation frequency in the generation of PA signal is in this case wavelength dependent ($\nu_{PA} \propto \nu_{IR}$). The FTIR plot of Fig. 2b thus includes a contribution that reflects the dependency of the PA response on the modulation frequency as well as on the absorber material. (Note that the laser measurements of Fig. 2b do not suffer from such an artifact because the laser modulation frequency was kept constant 40 Hz throughout the measurement). This and other features of the FTIR measurements are described in detail in Ref. [27].

Data analysis

The FFT spectrum represents the power spectrum of the PA signal, and the FFT signal peak (Fig. 1b) gives the PA signal reading. This is the value plotted *e.g.* in Fig. 3a as a function of the optical power measured by a reference detector. A linear least squares fit to the measured PA signal vs. the incident optical power gives the PA conversion factor $C_{PA}$ in arbitrary units/W, see Fig. 3a. The normalized noise-equivalent power $NEP = \frac{P_\sigma}{\sqrt{\Delta f}} = \frac{\sigma}{C_{PA}\sqrt{\Delta f}}$ is expressed in units W/√Hz. Here, σ is the 1σ root-mean-square noise of the FFT power spectrum at the modulation frequency and $P_\sigma$ is the corresponding noise-equivalent optical power (SNR = 1). The NEP is normalized by the square root of the (electrical) measurement bandwidth $\Delta f = 2/\tau$, where $\tau$ is the averaging time, *i.e.* the time constant of the FFT filter. In the measurements reported here, we used $\tau = 1.57$ s unless otherwise mentioned. The specific detectivity is calculated from $D^* = \sqrt{A}/NEP$, where $A = \pi (0.5 \text{ cm})^2$ is the area of the PA detector's absorber. The dependence of NEP on the averaging time $\tau$ was estimated by Allan deviation analysis[36] of the PA signal that was continuously recorded for 66 hours with $\tau = 1.57$ s. The calculation of the Allan standard deviation $NEP(\tau)$ presented in Fig. 3d was performed using *AlaVar 5.2* freeware tool. This analysis was done without any laser light (the detector seeing just the thermal background), which gives the noise level of the measurement at room temperature and thus represents the detector $NEP(\tau)$. Note that the definition of linear dynamic range used in this paper is $LDR = 10 \log(P_{max}/P_{min})$, where $P_{max}$ and $P_{min}$ are, respectively, the maximum and minimum measurable optical powers within the detector's linear range. For the lower limit we have used $P_{min} = NEP(\Delta f = 1 \text{ Hz}) = 5.6$ nW. (For photodetectors, definition $LDR = 20 \log(P_{max}/P_{min})$ is often used).

**Acknowledgements**

The work was funded by the Academy of Finland (Project numbers 326444 and 314364) and by the Academy of Finland Flagship Programme, Photonics Research and Innovation (PREIN), decision number: 320167.

**Author contributions**

M.V. and T.H. conceived the idea. J.R., M.V., J.U., T.H., T.L. and E.I. planned and discussed the experiments. J.R. carried out the experiments and data analysis with support from J.U., M.V. and S.S. The 14.85-µm quantum cascade laser needed in the work was designed and fabricated by R.T. and A.B. The manuscript was mainly written by J.R. and M.V. All authors reviewed the manuscript and provided editorial input. M.V., E.I. and T.L. supervised the project, M.V. and E.I. acquired funding.

**Data availability**

The data that support the findings of this study are available from the corresponding authors upon reasonable request.

**Conflict of interest**

The authors declare that they have no conflict of interest.